\documentclass[11pt]{amsart}
\usepackage{geometry}                
\usepackage{graphicx}
\usepackage{amssymb}
\usepackage{epstopdf}
\usepackage{array}
\usepackage{subfigure}
\usepackage{url}
\usepackage{hyperref}
\usepackage{xspace}
\usepackage{lscape}
\usepackage{isorot}

\usepackage{color}

\definecolor{gray}{rgb}{0.7,0.7,0.7}

\usepackage{supertabular}
\usepackage{hhline}
\newcommand\arraybslash{\let\\\@arraycr}
\makeatother
\title[Towards plant wire]{Towards plant wires}
\author[Adamatzky]{{\bf Andrew Adamatzky}\\ \vspace{0.1cm}  \\ Unconventional Computing Centre, \\ University of the West of England,\\ Bristol, BS16 1QY, United Kingdom}
\address[Adamatzky]{University of the West of England, Bristol, BS 16 1QY, United Kingdom}

\begin{document}

\maketitle

\begin{abstract}
In experimental laboratory studies we evaluate a possibility of making electrical wires from living plants. 
In scoping experiments we use lettuce seedlings as a prototype model of a plant wire.  We approximate an electrical potential transfer function by applying direct current voltage to the lettuce seedlings and recording output voltage.  We analyse oscillation frequencies of the output potential and assess noise immunity of the plant wires. Our findings will be used in future designs of self-growing wetware circuits and devices, and integration of  plant-based electronic components into future and emergent bio-hybrid systems.  

\end{abstract}

\section{Introduction}

Since its inception in 1980s the field of unconventional computation~\cite{calude_1998} became dominated by theoretical research, including quantum computation, membrane computing and dynamical-systems  computing.  Just a few experimental laboratory prototypes are designed so far~\cite{adamatzky_teuscher, teuscher_adamatzky}, e.g.   chemical reaction--diffusion processors~\cite{adamatzky_2005}, extended analog computers~\cite{mills_2008},  micro-fluidic circuits~\cite{fuerstman_2003}, gas-discharge systems~\cite{reyes_2002},  
chemo-tactic droplets~\cite{lagzi_2010}, enzyme-based logical circuits~\cite{katz_2010, privman_2009},  crystallization computers~\cite{hotice}, geometrically constrained chemical computers~\cite{sielewiesiuk_2001, motoike_2003, gorecki_2009, yoshikawa_2009, gorecki_2006a}, molecular logical gates and circuits~\cite{stojanovic_2002,mcdonald_2006}. In contrast, there are hundreds if not thousands of papers published on quantum computation, membrane computing  and artificial immune systems. Such a weak representation of laboratory experiments in the field of unconventional computation may be due to technical difficulties and costs of prototyping. 

In last few years we have developed a concept, architectures and experimental laboratory prototypes of living and hybrid computers made of slime mould \emph{Physarum polycephalum}, see e.g. \cite{adamatzky_physarummachines, adamatzky_alife}. In course of our studies of the slime mould's computational properties and designs of Physarum chips we found that the slime mould based sensors and processor are very fragile, highly dependent on 
environmental conditions and somewhat difficult to control and constrain.   Thus we have started to look for living substrates that could complement, or even become an alternative to, computing devices made of Physarum's protoplasmic tubes. Plant roots got our attention because they are mobile, growing, adaptive 
intelligent units~\cite{Brenner_2006,baluska_2009, baluska_2010, ciczak_2012, Burbach_2012}, similar in their behaviour to active growing zones of \emph{Physarum polycephalum}. Plants are, in general, more robust and resilient, 
 less dependent on environmental conditions and can survive in a hostile environment of bio-hybrid electronic devices longer than slime moulds do. 
 
 A prospective computing device made from living plant would perform computation by means of electrical charge propagation, by travelling waves of mechanical deformation of immobile structures, and by physical propagation  of structures. Conductive pathways would be a key component of any plant-based electrical circuit. Thus to achieve a 
 realistic assessment of the potential of the living plants based hybrid processing technology we must evaluate the feasibility of living plant wires in a proof-of-concept device. Previous findings on of organic wires and using living substrates to grow  conductive pathways: self-assembling molecular wires~\cite{wang_2006,paul_1998}, DNA wires~\cite{beratan_1997}, electron transfer pathways in biological systems~\cite{katz_2005}, live bacteria templates for conductive  pathways~\cite{berry_2005},  bio-wires with cardiac tissues~\cite{cingolani_2012}, golden wires with templates of fungi~\cite{sabah_2012}. We are not aware of any published results related to evaluation of living plants, or their parts, as electrical wires. There are substantial experimental findings on electrical impedance of plants however 
 mostly related to evaluation of the plants physiological state via their impedance characteristics, see 
 e.g.~\cite{zhang_1993, inaba_1995, mancuso_1997, mancuso_2000}.

The paper is structured as follows. We describe experimental techniques in Sect.~\ref{methods}. Section~\ref{results}
presents results on resistance of lettuce seedlings, approximation of direct current potential transfer function and immunity of plant wires to noise. Outcomes of the research and future directions of study are discussed in Sect.~\ref{discussion}.

\section{Methods}
\label{methods}

\begin{figure}[!tbp]
\centering
\subfigure[]{\includegraphics[width=0.5\textwidth]{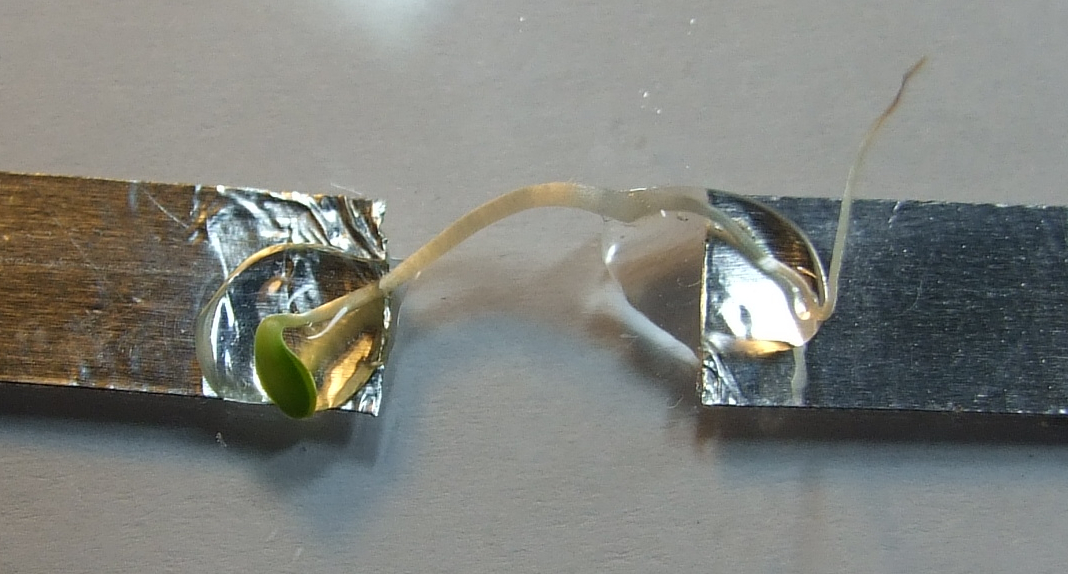}}
\subfigure[]{\includegraphics[width=0.25\textwidth]{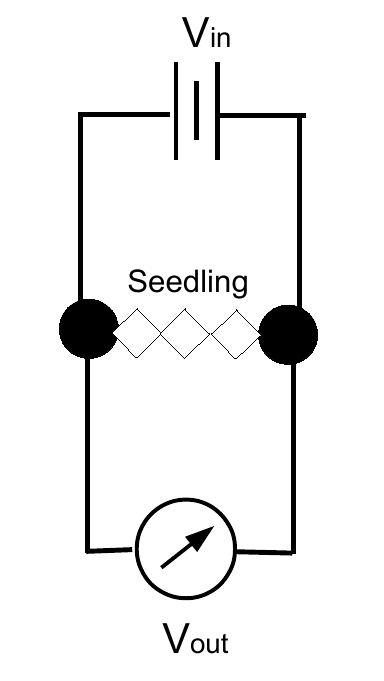}}
\caption{Experimental setup. 
(a)~A photo of lettuce seedling resting on aluminium electrodes in drops of distilled water.
(b)~A scheme of the experimental measurement of  potential transfer function.}
\label{experimental}
\end{figure}

We experimented with lettuce (\emph{Lactuca sativa}) seedlings 3-4 days after germination.
Electrodes were made of  a conductive aluminium foil, 0.07~mm thick,  8~mm wide, 50~mm (inclusive part protruding outside Petri dish) long. Distance between proximal sites of electrodes is always 10~mm. In each experiment an undamaged lettuce seedling was placed onto electrodes in drops of distilled water (Fig.~\ref{experimental}a).

In each experiment a resistance and potential (not at the same time indeed) were measured during 10~min with four wires using Fluke 8846A precision voltmeter, test current 1$\pm$0.0013 $\mu$A. Direct current potential was applied using Gw Instek GPS-1850D laboratory DC power supply. A scheme of potential transfer function measurement is 
shown in Fig.~\ref{experimental}b.

Each session of measurements lasted exactly 10~min. For each measurement a new seedling was used. We measured resistance of 20 seedlings, and potential transfer function of 5 seedlings for each value of input, applied, potential: 2~V, 3~V, $\ldots$, 12V.

\section{Results}
\label{results}

\begin{figure}[!tbp]
\centering
\subfigure[]{\includegraphics[width=.8\textwidth]{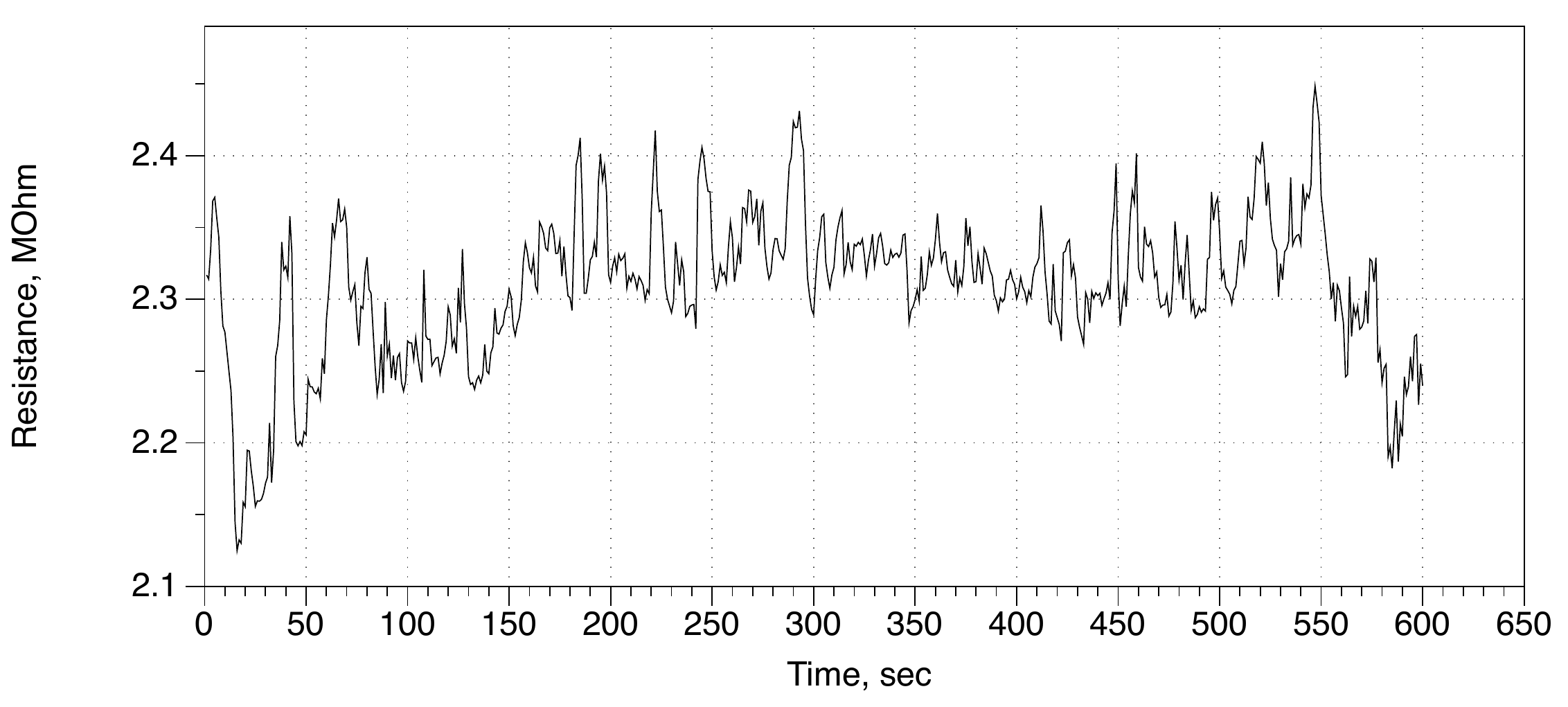}}
\subfigure[]{\includegraphics[width=.8\textwidth]{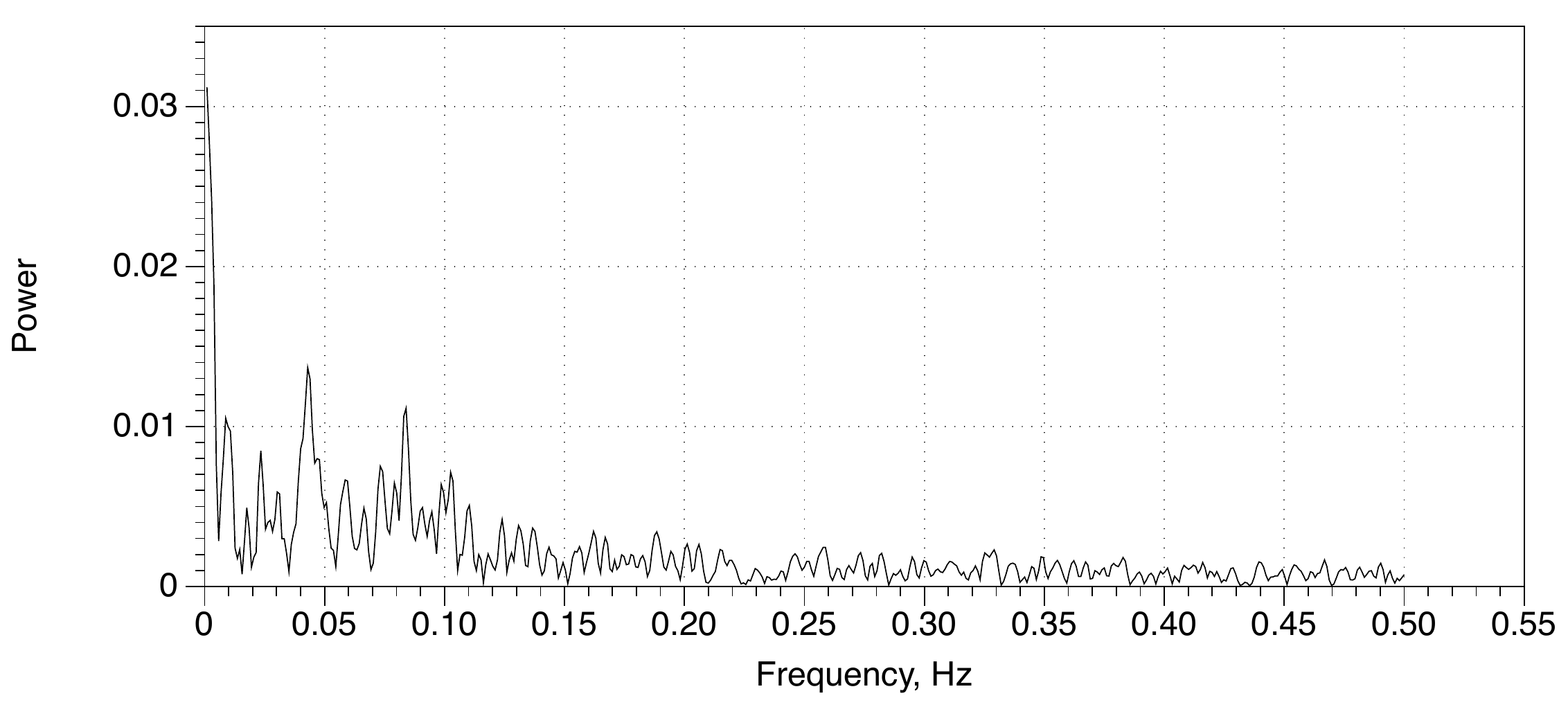}}
\caption{Lettuce resistance. (a)~Exemplar dynamics of resistance. (b)~Power frequency spectrum.}
\label{resistancegraph}
\end{figure}

Average resistance of lettuce seedling is 2.76~M$\Omega$ with standard deviation 0.18~M$\Omega$, and median 2.72~M$\Omega$.  The seedlings show oscillations of resistance (Fig.~\ref{resistancegraph}a). 
The oscillations are more likely is a combinations of dynamical changes in resistance in root, stem and leaves. Thus the oscillations recorded have no clearly visible periodicity. Three most dominating frequencies observed are 0.043~Hz, 0.084~Hz, and 0.009~Hz (Fig.~\ref{resistancegraph}b).

 \begin{figure}[!tbp]
\centering
\subfigure[]{
\begin{tabular}{l||l|l|l}
$V_{in}$, V	&	$V_{out}$, V	&	std($V_{out}$)	&	$\tilde{V_{out}}$, V  \\ \hline
2			&	0.82			&	0.011				&	0.82 \\
3			&	1.46			&	0.019				&	1.45\\
4			&	2.35			&	0.037				&	2.34\\
5			&	3.49			&	0.025				&	3.75\\
6			&	4.41			&	0.096				&	4.44 \\
7			&	5.06			&	0.092				&	5.08\\
8			&	6.5			&	0.044				&	6.50\\
9			&	6.67			&	0.016				&	6.68\\
10			&	8.46			&	0.115				&	8.46\\
11			&	8.81			&	0.096				&	8.84\\
12			&	10.00			&	0.120				&	10.02\\
\end{tabular}
}
\subfigure[]{\includegraphics[width=0.8\textwidth]{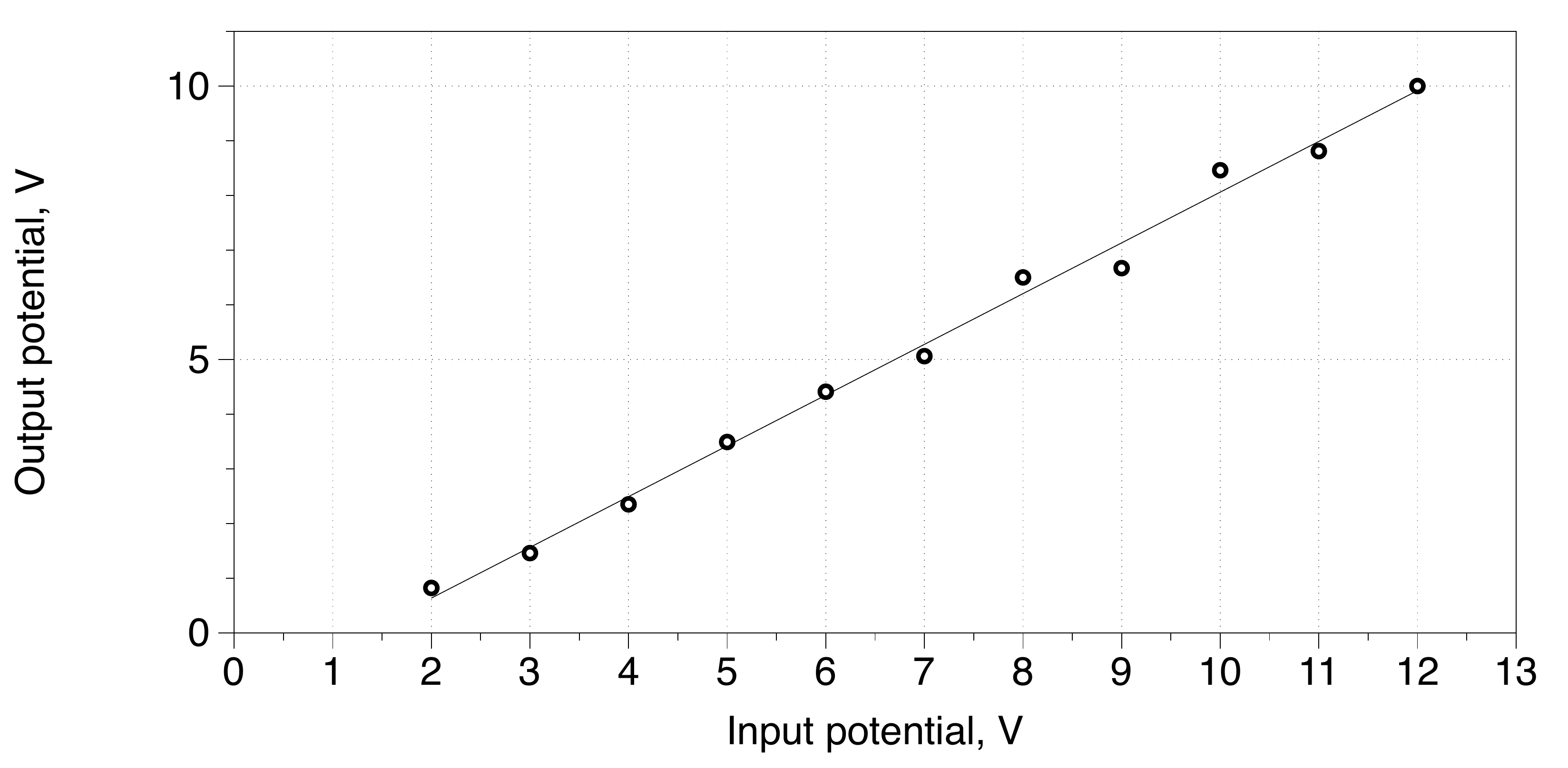}}
\caption{Potential transfer function implemented by lettuce seedling. 
(a)~A table of experimental measurements: for each value of input potential $V_{in}$ we show 
average output potential $V_{out}$, its standard deviation std($V_{out}$) and median output potential 
$\tilde{V_{out}}$.
(b)~Experimental plot of  a transfer function. 
}
\label{Vin2Vout}
\end{figure}

\begin{figure}[!tbp]
\centering
\subfigure[]{\includegraphics[width=0.8\textwidth]{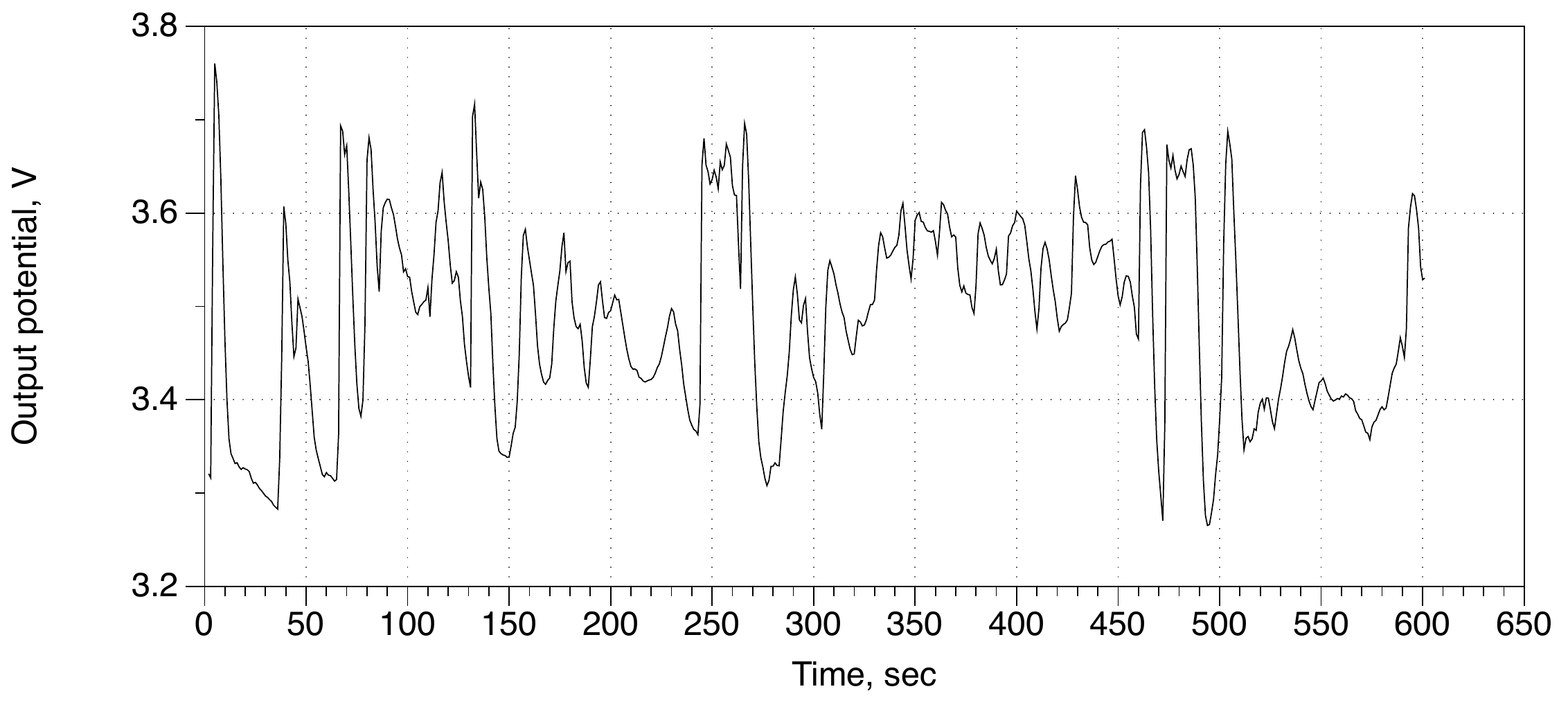}}
\subfigure[]{\includegraphics[width=0.8\textwidth]{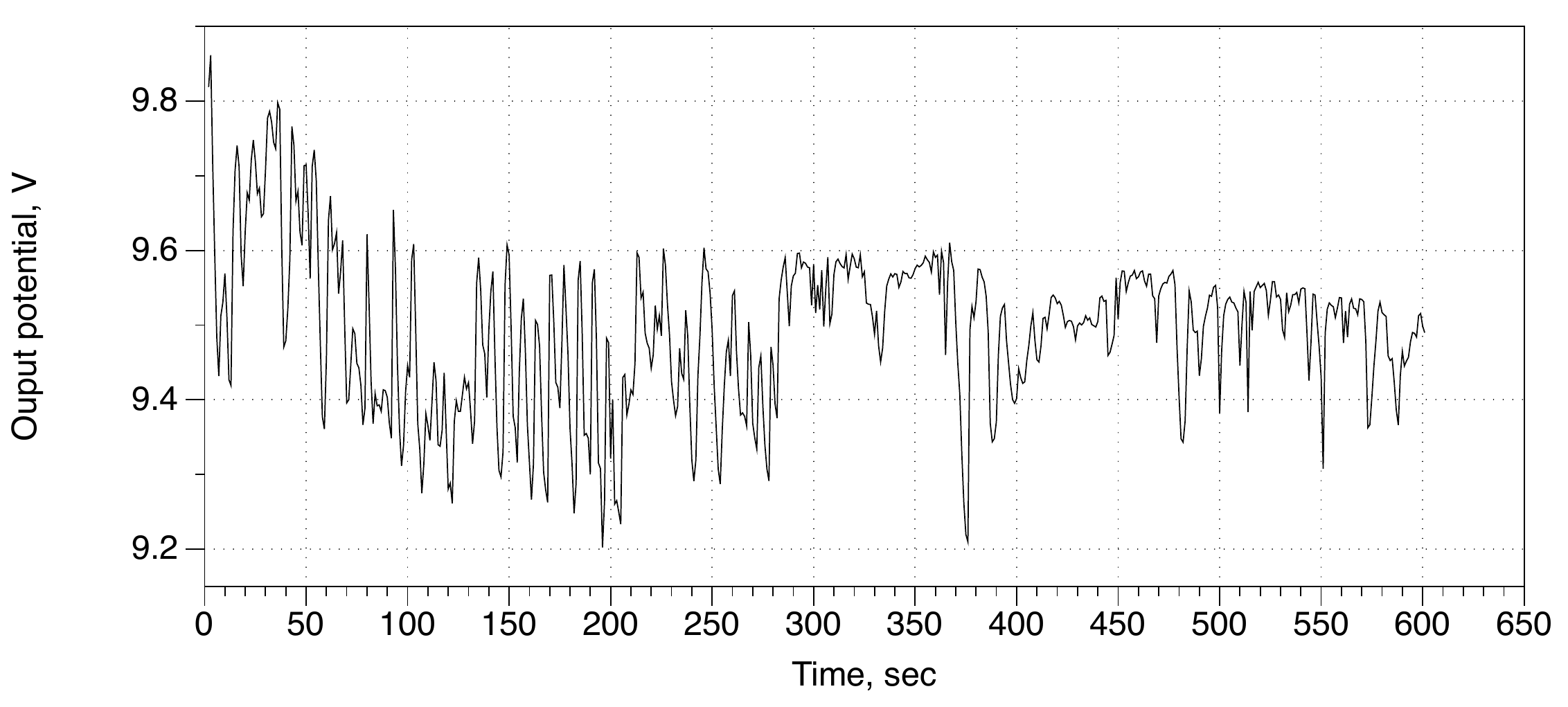}}
\caption{Dynamics of output potential for input potential 5~V~(a) and 12~V~(b). 
}
\label{examplesPotential}
\end{figure}

A lettuce seedling acts as a potential divider.  Potential $V_{out}$ recorded on a lettuce wire is a fraction of a potential $V_{in}$ applied to the lettuce wire.  Results of 60 experiments, 5 experiments for each value of input potential, are summarised in Fig.~\ref{Vin2Vout}a and illustrated in Fig.~\ref{Vin2Vout}b.  The transfer function is linear (Fig.~\ref{Vin2Vout}b) $V_{out}$ = 0.928$\cdot$$V_{in}$ - 1.221,   subject to oscillations of output potential (Fig.~\ref{examplesPotential}) caused by fluctuations of impedance (Fig.~\ref{resistancegraph}a).  

\begin{figure}[!tbp]
\centering
\subfigure[]{
\begin{tabular}{l||l|l|l|l}
$V_{in}$, V	&		&	average, Hz	&	st. deviation	&	average, sec	\\ \hline
2	&	$f_1$	&	0.0117	&	0.00626	&	85.65	\\
	&	$f_2$	&	0.0310	&	0.03535	&	32.24	\\
3	&	$f_1$	&	0.0320	&	0.03167	&	31.21	\\
	&	$f_2$	&	0.0295	&	0.01056	&	33.9	\\
4	&	$f_1$	&	0.0088	&	0.00328	&	113.96	\\
	&	$f_2$	&	0.0208	&	0.00700	&	48.13	\\
5	&	$f_1$	&	0.0176	&	0.02172	&	56.92	\\
	&	$f_2$	&	0.0149	&	0.01236	&	67.1	\\
6	&	$f_1$	&	0.0190	&	0.01147	&	52.49	\\
	&	$f_2$	&	0.0163	&	0.00578	&	61.26	\\
7	&	$f_1$	&	0.0078	&	0.00082	&	128.21	\\
	&	$f_2$	&	0.0247	&	0.01680	&	40.53	\\
8	&	$f_1$	&	0.0139	&	0.00922	&	72.05	\\
	&	$f_2$	&	0.0180	&	0.00697	&	55.68	\\
9	&	$f_1$	&	0.0205	&	0.01690	&	48.78	\\
	&	$f_2$	&	0.0260	&	0.02271	&	38.49	\\
10	&	$f_1$	&	0.0120	&	0.00374	&	83.51	\\
	&	$f_2$	&	0.0251	&	0.01915	&	39.76	\\
11	&	$f_1$	&	0.0062	&	0.00346	&	161.29	\\
	&	$f_2$	&	0.0241	&	0.01290	&	41.49	\\
12	&	$f_1$	&	0.0299	&	0.03295	&	33.47	\\
	&	$f_2$	&	0.0385	&	0.01487	&	25.99	\\
\end{tabular}
}
\subfigure[]{\includegraphics[width=0.9\textwidth]{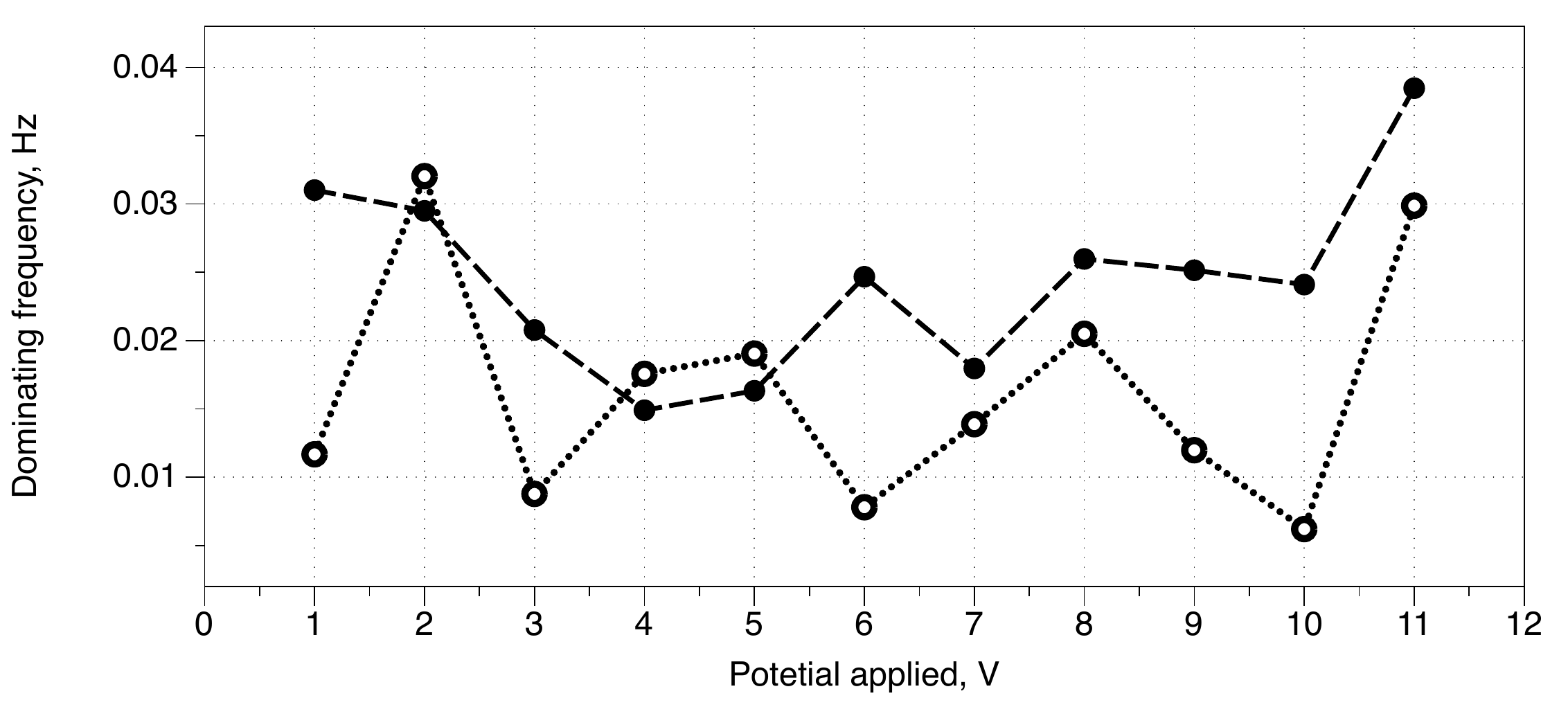}}
\caption{Two dominating frequencies of output potential oscillations for input potential applied 2~V, $\ldots$, 12~V.
(a)~Table of first $f_1$ and second $f_2$ dominating frequencies, their values in Herz and seconds.
(c)~Average dominating frequencies for $V_{in}$= 2~V, $\ldots$, 12~V: $f_1$ is shown by circles,
$f_2$ is shown by discs.}
\label{Frequency_vs_Vinp}
\end{figure}

 Resistance of lettuce seedlings changes during recording. 
 This causes oscillations of output potential (Fig.~\ref{examplesPotential}). Values of 
 two dominating frequencies, calculated by  FFT,  are shown in Fig.~\ref{Frequency_vs_Vinp}a. For most values of input potential dominating frequencies of output potential oscillations stay in a range between 0.05~Hz and 0.3~Hz 
 (Fig.~\ref{Frequency_vs_Vinp}b). There is no correlation between frequency of output potential oscillations 
 and values of input potential applies.
 
Output  potential oscillates in lower frequencies than resistance. For values of input potential studied
frequencies of output potential oscillations lie in a range between 0.0062~Hz and 0.038~Hz (with average 0.0221~Hz and median 0.0197~Hz over all  values of input potential).    

\begin{figure}[!tbp]
\centering
\subfigure[]{\includegraphics[width=0.8\textwidth]{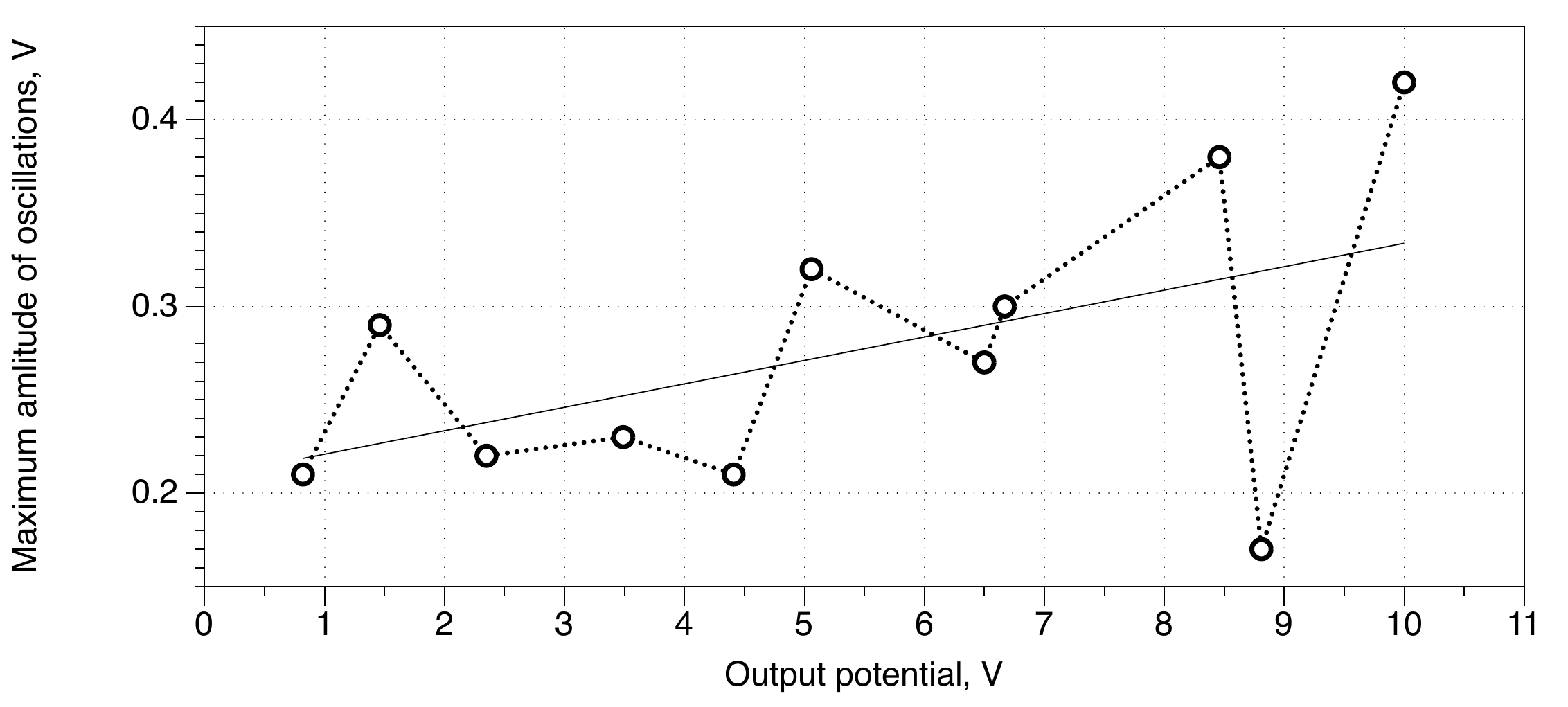}}
\subfigure[]{\includegraphics[width=0.8\textwidth]{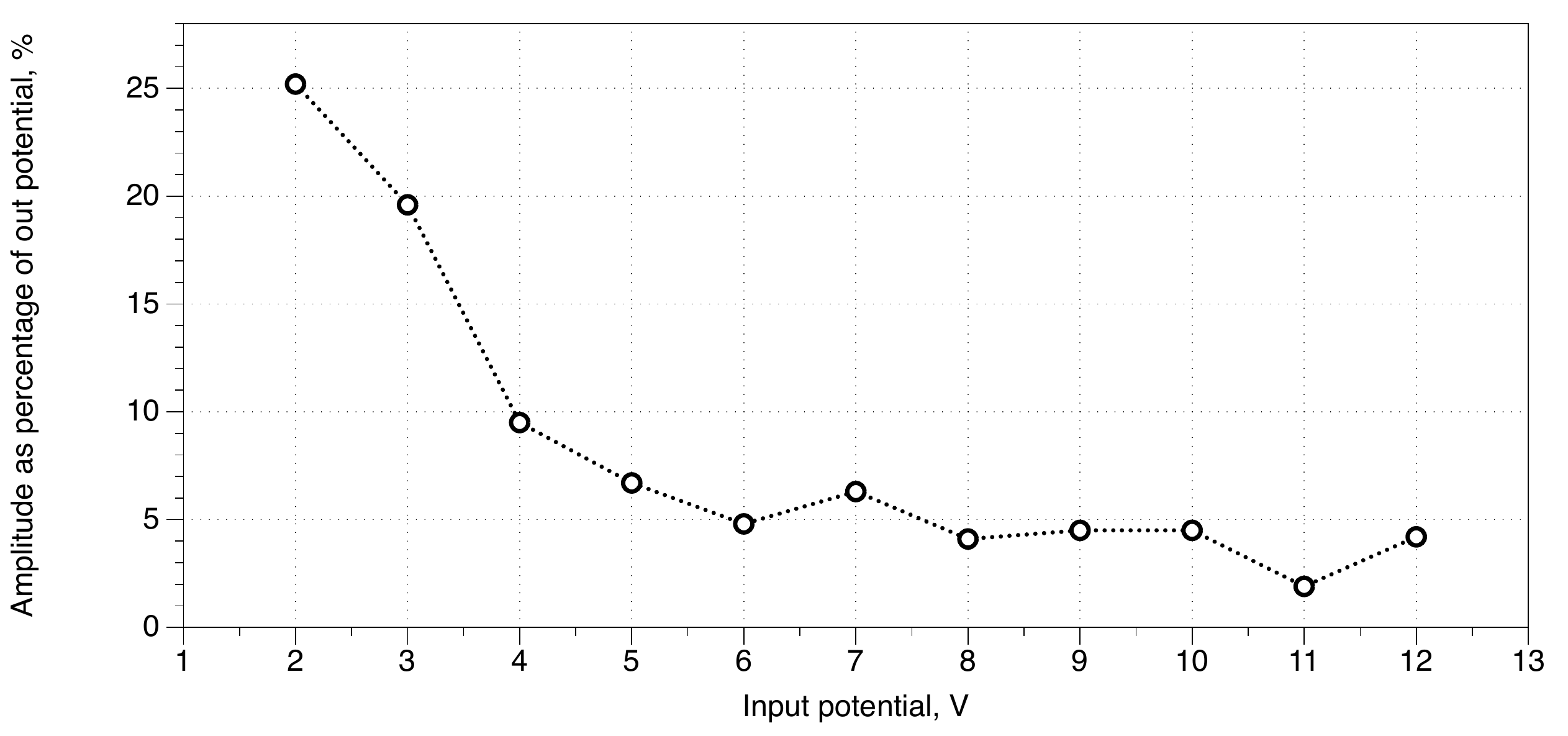}}
\caption{Maximum amplitudes of output potential oscillations.
(a)~Maximum amplitude of output potential oscillations, averaged over trials for each value of input potential, versus average output potential. Data points derived from laboratory experiments are shown by circle. Dotted line acts as an eye guide. Solid line is a linear fit 0.208 + 0.0126$\cdot$$V_{out}$. 
(b)~Maximum amplitude as a percentage of average output potential. Experimental data are shown by circles. Dotted line guides eye. 
}
\label{Vout_vs_MaxAmplitude}
\end{figure}

We studied lettuce seedling as a prototypes of plant wires. The wires are noisy because the output potential oscillates (Fig.~\ref{examplesPotential}). Level of noise can be evaluated via maximum amplitudes of oscillations. The maximum amplitudes of output potential oscillations are characterised in Fig.~\ref{Vout_vs_MaxAmplitude}a.  The maximum amplitudes stay in a range 0.15~V to 0.45~V and just slightly increase (rough linear approximation 0.208 + 0.0126$\cdot$$V_{out}$) with increase of a potential. The maximum amplitude makes 25\% of output potential for input potential 2V, then decreases to 19.6\% (3V), 9.5\% (4V), stays between 5\% and 7\% for 5-7V applied potential, and then decreases to 4\%--4.5\% for input potential 8-12V (Fig.~\ref{Vout_vs_MaxAmplitude}).

\section{Discussions}
\label{discussion}

In laboratory experiments with lettuce seedlings we found that they implement a linear transfer function of input potential to output potential. Roughly an output potential is 1.5--2~V less than an input potential, thus e.g. by applying
12~V potential we get 10~V output potential. Resistance of 3-4 day lettuce seedling is about 2.76~M$\Omega$. This is much higher than resistance of conventional conductors yet relatively low comparing to resistance of other living creatures~\cite{geddes_1967}, e.g. a resistance of 10~mm long protoplasmic tube of \emph{P. polycephalum} is
near 3~M$\Omega$~\cite{adamatzky_oscillator}. 

The lettuce resistance oscillates. The dominating frequencies of lettuce seedlings resistance oscillation roughly corresponds to periods of oscillation 23~sec, 12~sec and 111~sec. These are compatible with periods of resistance oscillations recorded in our experiments with protoplasmic tubes of acellular slime mould \emph{P. polycephalum}~\cite{adamatzky_oscillator}: 73~sec is an average period of oscillation.  Resistance oscillation in \emph{P. polycephalum} are caused by  peristaltic oscillations~\cite{sun_2009, adamatzky_jones_2011} and cytoplasmic flow in the tubes~\cite{adamatzky_schubert}. The peristaltic and flow  are governed by calcium waves travelling along the tubes where  a calcium ion flux through membrane triggers oscillators responsible for dynamic of contractile activity~\cite{meyer_1979,fingerle_1982}. Thus we can speculate that there are  types of cytoplasmic flow in lettuce seedlings which might be reflected in changes of volume of leaves, roots or stem that are reflected in resistance changes.

Lettuce seedling is somewhat a noisy wire. Its output potential oscillates. However, maximum amplitude of oscillations never exceeds 4\% of an input potential for potential applied over 8~V.  For example, a line held near 10~V (output voltage) has 9~V of noise immunity.

\begin{figure}[!tbp]
\centering
\subfigure[]{\includegraphics[width=0.49\textwidth]{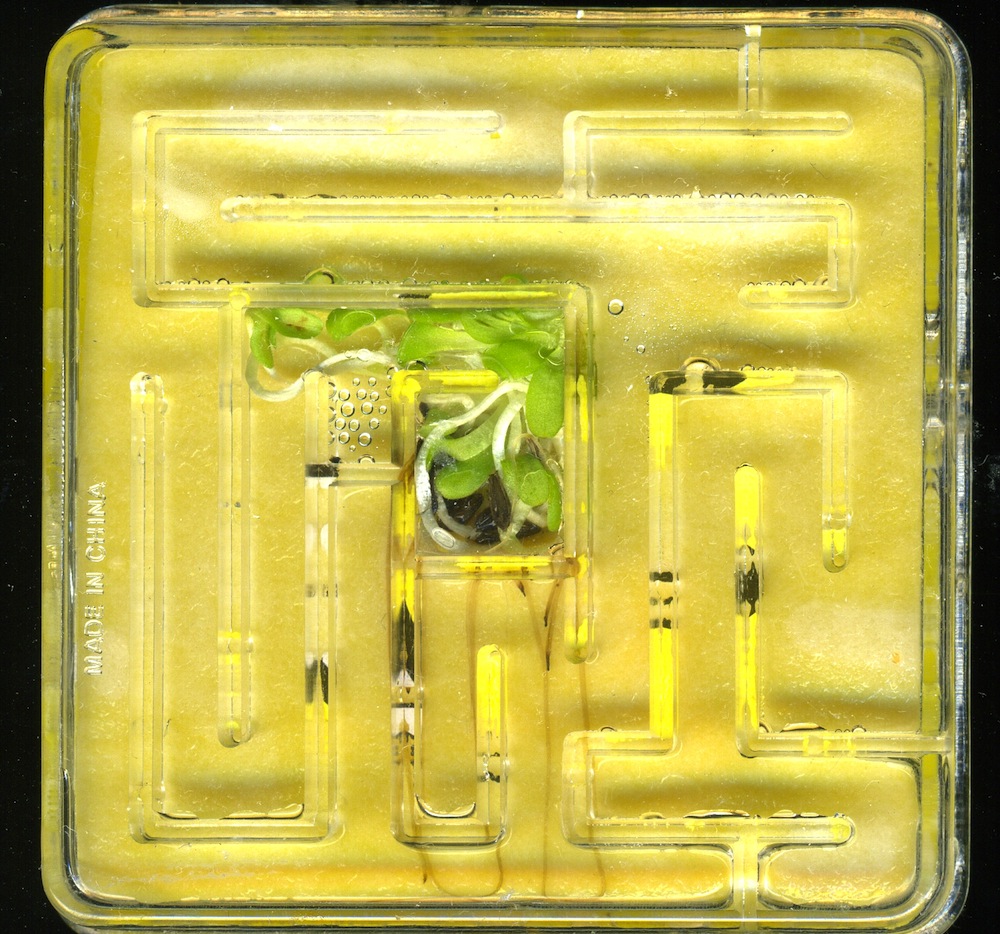}}
\subfigure[]{\includegraphics[width=0.49\textwidth]{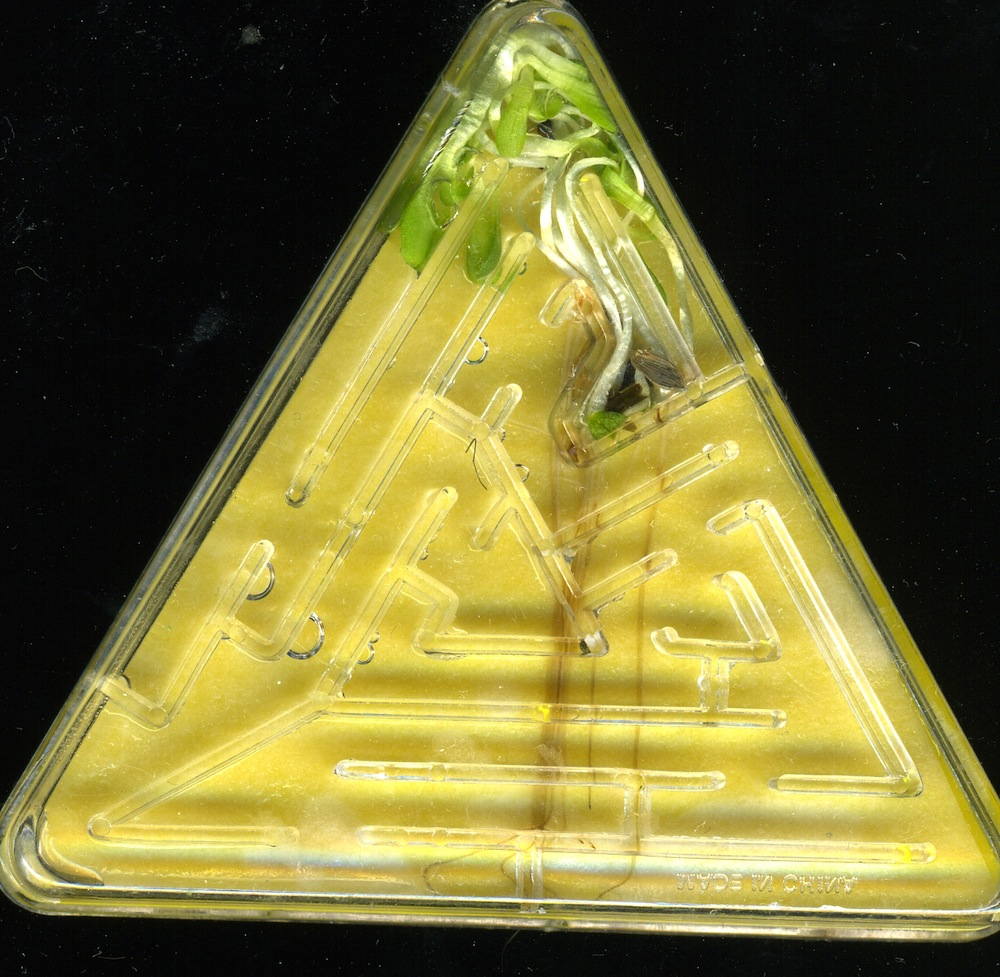}}
\caption{Scoping experiments on routing plant roots in labyrinths.}
\label{maze}
\end{figure}

To incorporate plant wires into bio-hybrid self-growing circuits we must develop techniques for reliable routing 
of the plant roots between living and silicon components of the circuits. A first step would be to find a way of
navigating plant roots in labyrinths. We have undertook few scoping experiments, see two illustrations in 
Fig.~\ref{maze}, yet no reached no conclusive results. When seeds are placed in or near a central chamber of a
labyrinth their routes somewhat grow towards exit of the labyrinth. However, they often become stuck midway and do not actually reach the exit.   The inconclusive results are possibly due to the fact that we used gravity as  the only guiding force to navigate the routes. In their recent paper Yokawa and colleagues~\cite{yokawa_2013} demonstrated
that by employing chemotaxis and using volatiles it is possible to navigate the roots in simple binary mazes. Their results 
gives us a hope it could be possible to route plant wires in the same efficient manner as slime mould based wires are routed~\cite{adamatzky_physarummaze, adamatzky_wires}. Another line of future research would be to investigate suitability of live plants to be incorporated as sensors in bio-hybrid electronic circuits. Thermistors could be the first port of call because there is an experimental evidence that plants resistance is changed during cooling~\cite{mancuso_2000}.

\end{document}